\documentclass[preprint]{aastex}
\usepackage{natbib, aas_macros}
\shorttitle{Photo-dissociation of molecules in Cygnus-X}
\shortauthors{Yamagishi, M. et al.}

\begin{document}

\title{Nobeyama 45m Cygnus-X CO survey I: photodissociation of molecules revealed by the unbiased large-scale CN and C$^{18}$O maps}

\author{M. Yamagishi\altaffilmark{1}, A. Nishimura\altaffilmark{2}, S. Fujita\altaffilmark{2}, T. Takekoshi\altaffilmark{3}, M. Matsuo\altaffilmark{4}, T. Minamidani\altaffilmark{4,5}, K. Taniguchi\altaffilmark{4,5}, K. Tokuda\altaffilmark{6,7}, Y. Shimajiri\altaffilmark{8}}
\email{yamagish@ir.isas.jaxa.jp}

\altaffiltext{1}{Institute of Space and Astronautical Science, Japan Aerospace Exploration Agency, Chuo-ku, Sagamihara 252-5210, Japan}
\altaffiltext{2}{Graduate School of Science, Nagoya University, Furo-cho, Chikusa-ku, Nagoya 464-8602, Japan}
\altaffiltext{3}{Institute of Astronomy, The University of Tokyo, 2-21-1 Osawa, Mitaka, Tokyo 181-0015, Japan}
\altaffiltext{4}{Nobeyama Radio Observatory, National Astronomical Observatory of Japan (NAOJ), National Institutes of Natural Sciences (NINS), 462-2, Nobeyama, Minamimaki, Minamisaku, Nagano 384-1305, Japan}
\altaffiltext{5}{Department of Astronomical Science, School of Physical Science, SOKENDAI (The Graduate University for Advanced Studies), 2-21-1, Osawa, Mitaka, Tokyo 181-8588, Japan}
\altaffiltext{6}{National Astronomical Observatory of Japan, National Institutes of Natural Sciences, 2-21-1 Osawa, Mitaka, Tokyo 181-8588, Japan}
\altaffiltext{7}{Graduate School of Science, Osaka Prefecture University, 1-1 Gakuen-cho, Naka-ku, Sakai, Osaka 599-8531, Japan}
\altaffiltext{8}{Laboratoire AIM, CEA/DSM-CNRS-Universit\'e Paris Diderot, IRFU/Service d'Astrophysique, CEA Saclay, F-91191 Gif-sur-Yvette, France}

\begin{abstract}
We present an unbiased large-scale (9~deg$^2$) CN ($N$=1--0) and C$^{18}$O ($J$=1--0) survey of Cygnus-X conducted with the Nobeyama 45m Cygnus-X CO survey.
CN and C$^{18}$O are detected in various objects towards the Cygnus-X North and South (e.g., DR17, DR18, DR21, DR22, DR23, and W75N).
We find that CN/C$^{18}$O integrated intensity ratios are systematically different from region to region, and are especially enhanced in DR17 and DR18 which are irradiated by the nearby OB stars.
This result suggests that CN/C$^{18}$O ratios are enhanced via photodissociation reactions.
We investigate the relation between the CN/C$^{18}$O ratio and strength of the UV radiation field.
As a result, we find that CN/C$^{18}$O ratios correlate with the far-UV intensities, $G_0$.
We also find that CN/C$^{18}$O ratios decrease inside molecular clouds, where the interstellar UV radiation is reduced due to the interstellar dust extinction.
We conclude that the CN/C$^{18}$O ratio is controlled by the UV radiation, and is a good probe of photon-dominated regions.
\end{abstract}


%%-------------------------------------------------------------------------------------------------------------------------------------------------------
\section{Introduction}

A deep understanding of the interactions between the ultraviolet (UV) radiation and the interstellar medium (ISM) is essential to study evolutions of molecules in the Universe.
The most common sites for such interactions are PDRs (photon dominated regions), where photodissociation and photoionization reactions are the most relevant processes.
One of the most commonly used probes of PDRs in radio astronomy is CN (\citealt{Fuente93, Greaves96, Simon97, Fuente05}); one of the formation pathways leading to CN is photodissociation of HCN.
Based on their models, \citet{Boger05} showed that CN/HCN abundance ratios are high at the surface of a molecular cloud due to significant contribution of the interstellar UV radiation.
Another important probe is a rare isotope of CO (\citealt{Glassgold85, Yurimoto04}); UV radiation selectively dissociates rare CO isotopes more effectively than major CO isotopes because of the difference in the self shielding (\citealt{Lada94, Shimajiri14, Lin16}).
Until now, however, studies of PDR  probes are still limited, and there is no consensus about what the most reliable PDR probe is, because mapping observations of such weak lines are difficult even for high-mass star-forming regions.
Hence it is valuable to test possible PDR probes for a variety of high-mass star-forming regions.

Among many Galactic high-mass star-forming regions, Cygnus-X ($d$=1.4~kpc; \citealt{Rygl12}) is one of the best sites to study interactions between UV and the ISM.
Cygnus-X consists of nine OB associations (\citealt{Humphreys78}).
The most active and well-known one is Cygnus OB2 association which includes 169 OB stars (\citealt{Wright15}).
Cygnus-X is roughly divided into two regions: the Cygnus-X North and South.
Cygnus-X North shows many filamentary structures in infrared and radio.
Many young stellar objects (YSOs) are associated with Cygnus-X North (\citealt{Beerer10}).
By contrast, Cygnus-X South shows relatively diffuse structures, and the star-formation activity is less intense; \citet{Schneider06} showed that the average density and excitation temperature of CO in Cygnus-X North are higher than those in Cygnus-X South, and suggested that Cygnus-X North is a more active star-forming region.
Cygnus-X has been intensively observed in many studies, which are summarized by \citet{Reipurth08}.
In addition, recent large-area surveys of Cygnus-X have been performed by James Clerk Maxwell Telescope (JCMT; \citealt{Gottschalk12}), Five College Radio Astronomical Observatory (FCRAO; \citealt{Schneider10}), $Spitzer$ (\citealt{Hora07}), $Herschel$ (\citealt{Schneider16}), and the Balloon-borne Large Aperture Submillimeter Telescope (BLAST; \citealt{Roy11}).

In this paper, we present unbiased large-scale CN ($N$=1--0) and C$^{18}$O ($J$=1--0) maps of Cygnus-X, and test the availability of the CN/C$^{18}$O integrated intensity ratio as a PDR probe.
Since CN will be enhanced and C$^{18}$O will be selectively dissociated in PDRs, CN/C$^{18}$O ratios are expected to be sensitive to the UV radiation.
The present study is based on the Nobeyama 45m Cygnus-X CO survey (PI: A. Nishimura), where $^{12}$CO ($J$=1--0), $^{13}$CO ($J$=1--0), C$^{18}$O ($J$=1--0), and CN ($N$=1--0) were simultaneously observed for a 9 deg$^2$ area thanks to the large-scale mapping capability and the wide frequency coverage of the facility.
The simultaneous observations ensure reliability of the resultant CN and C$^{18}$O data.
The number of unbiased CO multi-line surveys for high-mass star-forming regions with high angular resolution is still limited (Orion: \citealt{Nishimura15, Shimajiri14}, Galactic plane: \citealt{Barnes15, Umemoto17}).
As for CN, such survey is even more limited (\citealt{Rodriguez-Franco98, Barnes15, Pety17}).
Therefore, our survey will provide a valuable data set not only for PDR studies but also for comparison to other surveys of Cygnus-X and for a general understanding of the ISM in massive star forming regions.
In this paper, we use the C$^{18}$O data for comparison to the CN data.
In addition, we supplementarily use the $^{13}$CO data for analyses of the CN and C$^{18}$O data.
Detailed analyses of the C$^{18}$O and $^{13}$CO data (e.g., properties of dense cores) will be presented in separate papers.

%%-------------------------------------------------------------------------------------------------------------------------------------------------------
\section{Observation and data reduction}

\begin{figure}
\epsscale{0.9}
\plotone{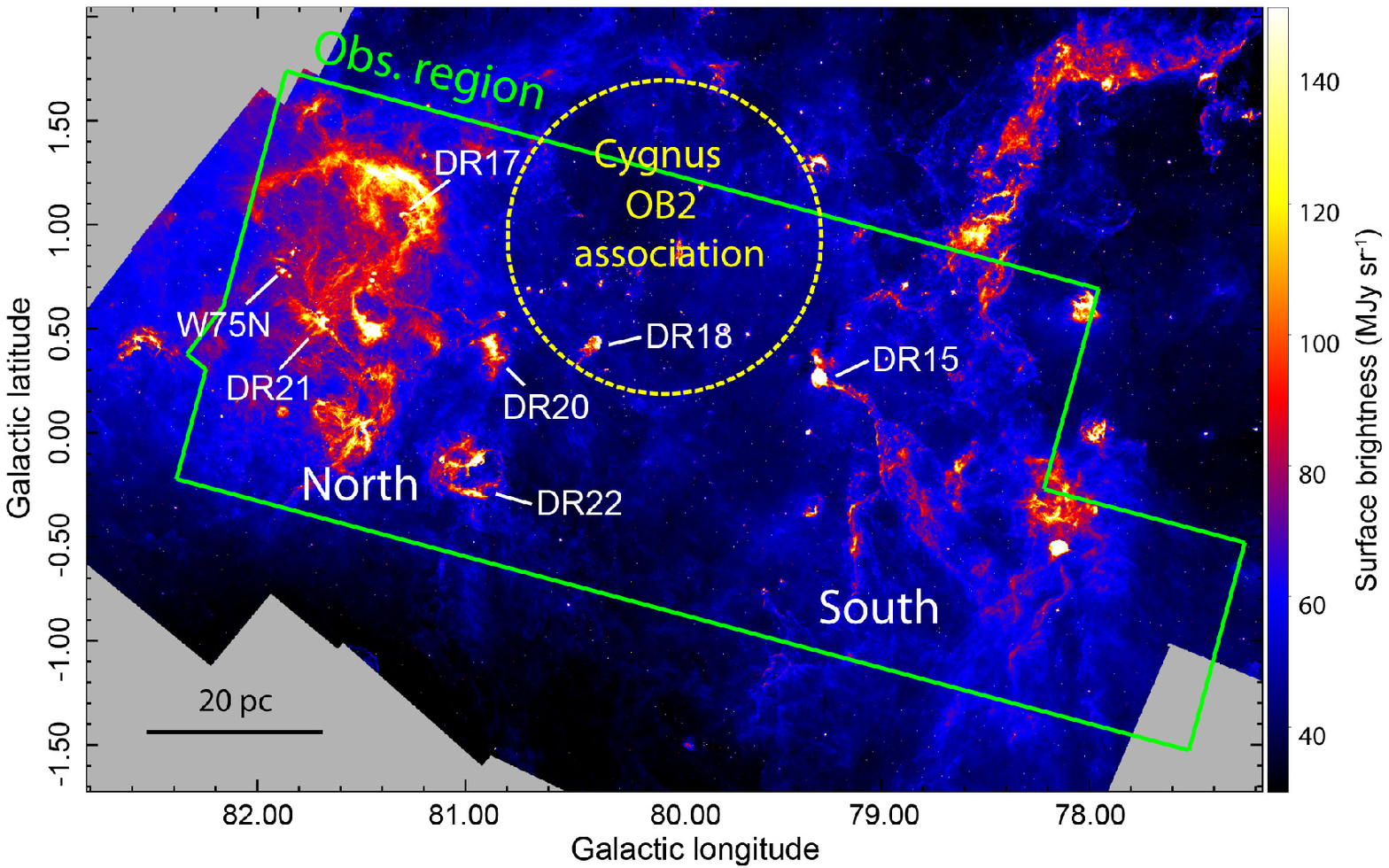}
\caption{Survey area overlaid on the $Spitzer$/IRAC 8~$\micron$ map taken by the $Spitzer$ Cygnus-X legacy survey (\citealt{Hora07}).}
\label{obsregion}
\end{figure}

The observations of Cygnus-X were carried out during 2016 January 13 and May 08 with the Nobeyama 45m/FOREST+SAM45.
The total observation time was 77 hrs.
The four-beam receiver FOREST (\citealt{Minamidani16a}) allows simultaneous multi-line on-the-fly mapping.
In the present survey, $^{12}$CO ($J$=1--0; 115.271202~GHz), $^{13}$CO ($J$=1--0; 110.201354~GHz), C$^{18}$O ($J$=1--0; 109.782176~GHz), and CN ($N$=1--0, $J$=3/2--1/2, $F$=5/2--3/2; 113.490982~GHz) were simultaneously observed in the single polarization mode.
The survey area is shown in Fig.~\ref{obsregion}, which covers both the Cygnus-X North and South as well as the Cygnus OB2 association.
The overall observed area is 9 deg$^2$, which was covered by mosaic sub-maps with $1^\circ\times1^\circ$.
The OTF scan parameters are the same as the FUGIN galactic plane survey (\citealt{Umemoto17}); the RX angle, scan spacing, scan length, scan speed, and sampling time are 9.$\arcsec$46, 8.$\arcsec$5, 3600$\arcsec$, 100$\arcsec$/sec, and 40 msec, respectively.
The typical system temperatures including atmosphere are 350~K, 200~K, 150~K, and 150~K for $^{12}$CO, CN, $^{13}$CO, and C$^{18}$O, respectively.
We performed multiple scans to achieve Nyquist sampling for each sub-map.
Pointing errors were corrected every 1.5 hr by observing a SiO maser source, AU Cyg.
As a result, the pointing accuracy of the telescope was kept to be $<5\arcsec$.

Data reduction was performed by using the NOSTAR software provided by the Nobeyama Radio Observatory.
For each map, we split data to each array, and subtracted baselines using a first-order polynomial.
The velocity ranges to subtract baselines are -150 -- -90 and 30 -- 100 km~s$^{-1}$ for CN, -150 -- -20 and 20 -- 100 km~s$^{-1}$ for C$^{18}$O, and -150 -- -60 and 25 -- 100 km~s$^{-1}$ for $^{13}$CO.
The amplitude of line intensities was calibrated based on the observational results of W51 and DR21.
The calibration method is the same as that applied in the FUGIN project (\citealt{Umemoto17}).  
We used a spatial grid of 22.$\arcsec$7 and a velocity grid of 0.5 km~s$^{-1}$ to obtain high signal-to-noise ratio maps.
Bessel-Gauss function was used for the convolution.
The effective angular resolution is 46$\arcsec$ (0.30~pc at $d$=1.4~kpc).
The typical noise levels of the final CN, C$^{18}$O, and $^{13}$CO data are 0.38~K, 0.26~K, and 0.27~K, respectively, in $T_{\mathrm{mb}}$ scale.
The final calibrated FITS cube of CN will be publicly released on the web\footnote{ https://cygnus45.github.io/}\footnote{http://www.nro.nao.ac.jp/\~{}nro45mrt/html/results/data.html}.
We also plan to release the FITS cubes of $^{12}$CO, $^{13}$CO, and C$^{18}$O after publishing the forthcoming science papers.

%%-------------------------------------------------------------------------------------------------------------------------------------------------------
\section{Results}

\begin{figure}
\epsscale{.5}
\plotone{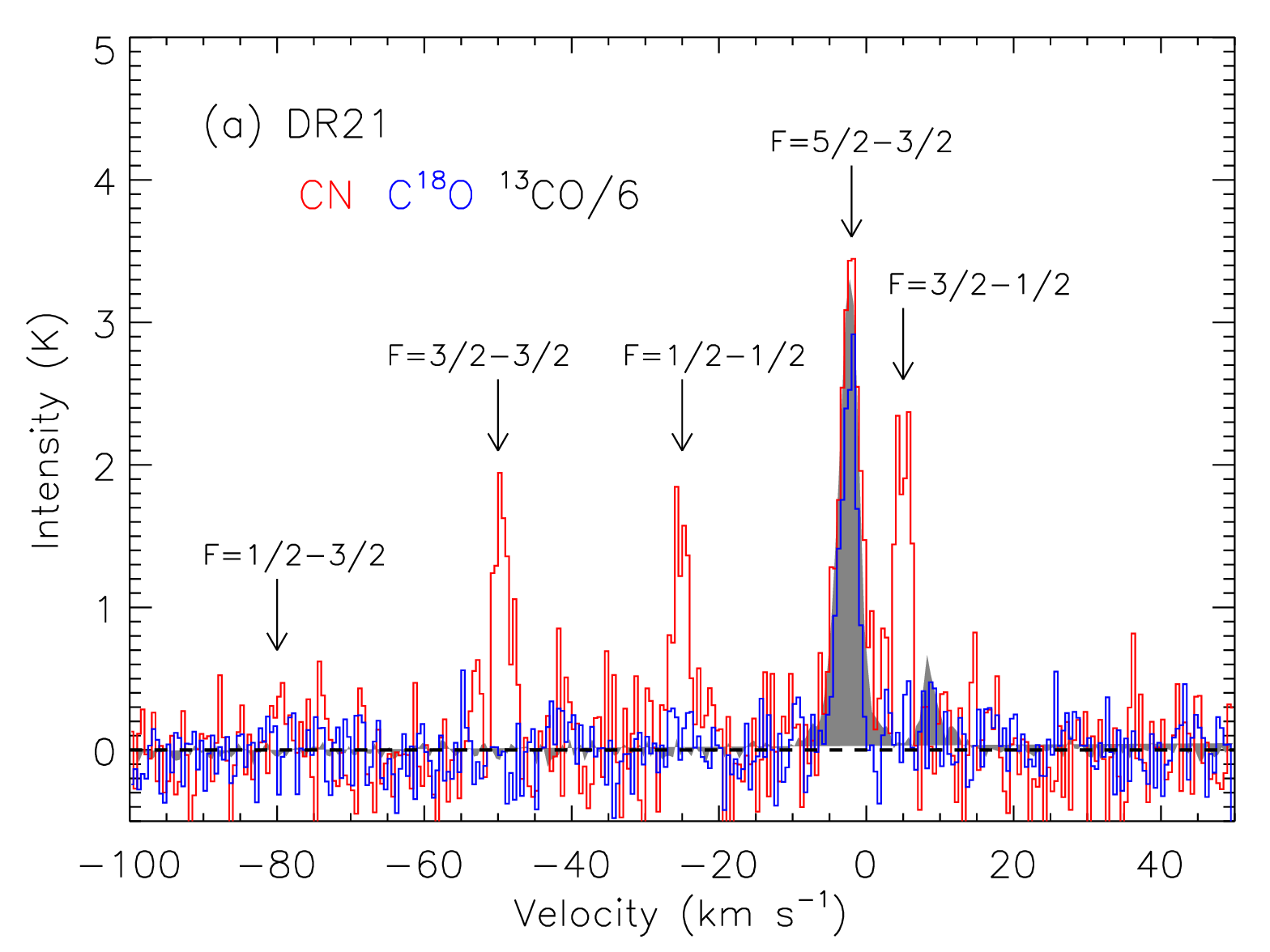}
\plotone{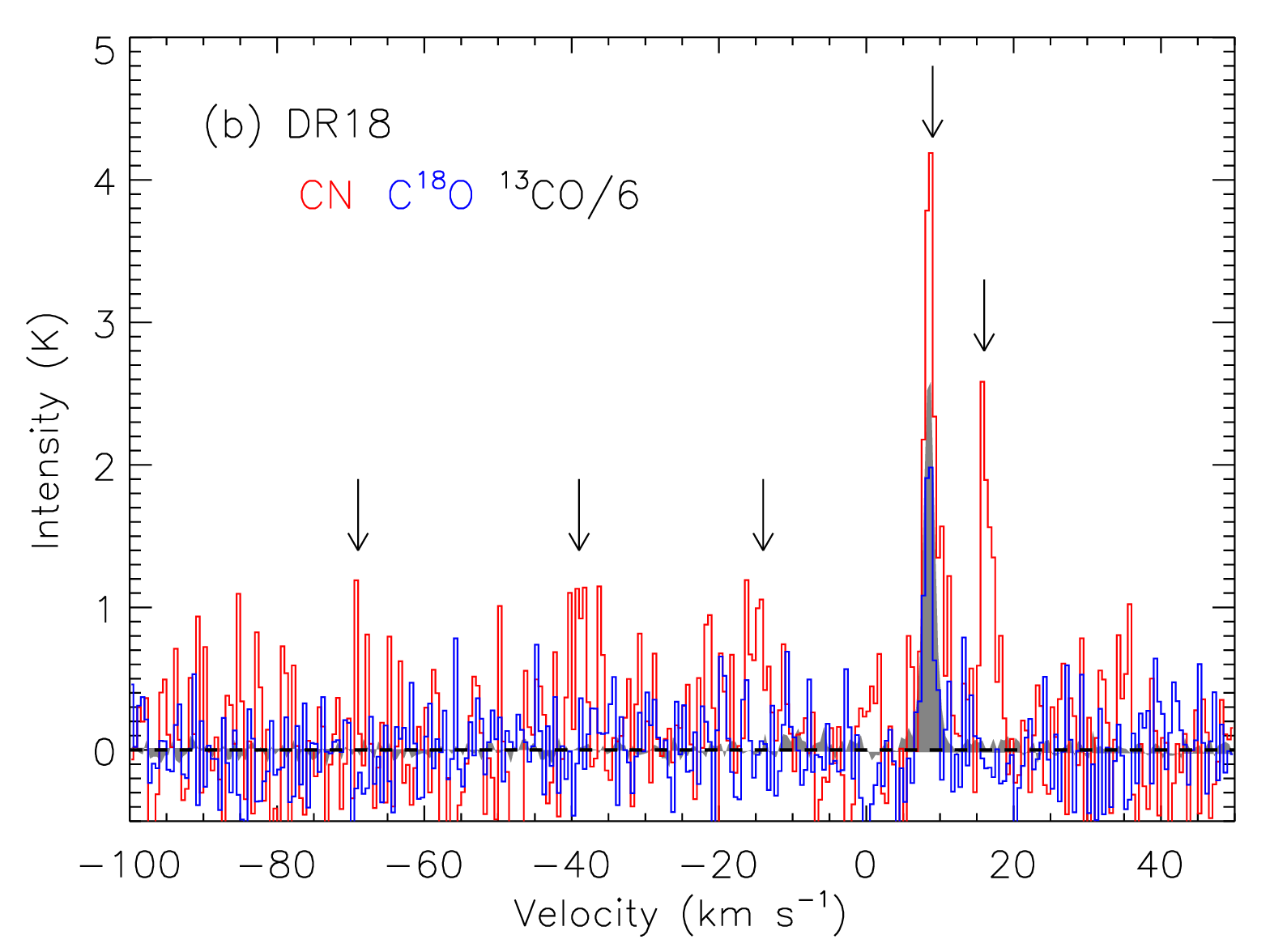}
\plotone{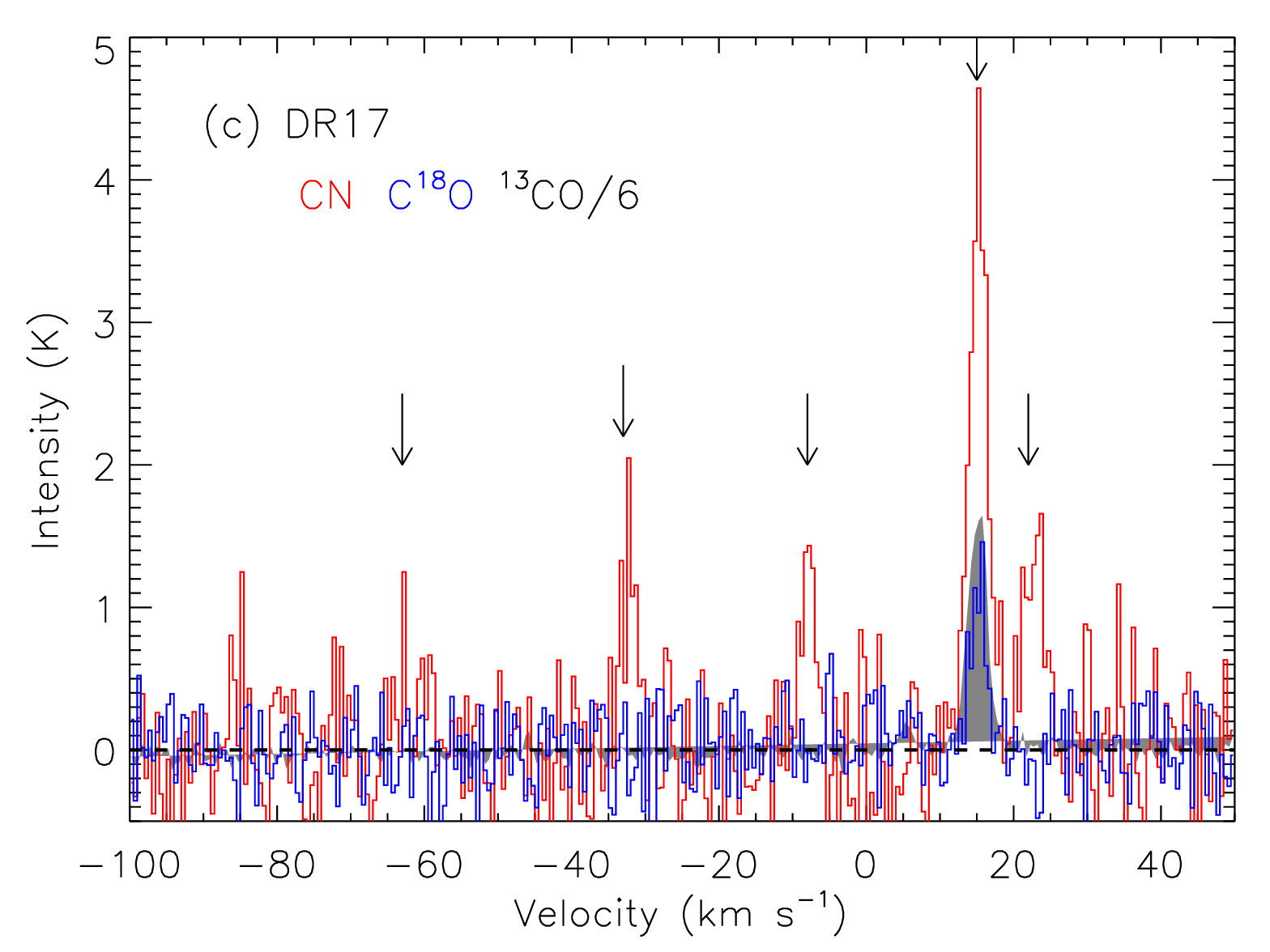}
\caption{Examples of CN (red), C$^{18}$O (blue), and $^{13}$CO (gray shade) spectra extracted from local peaks in (a) DR21 ($l$=81.681$^\circ$, $b$=0.546$^\circ$), (b) DR18 ($l$=80.369$^\circ$, $b$=0.451$^\circ$), and (c) DR17 ($l$=81.302$^\circ$, $b$=1.050$^\circ$). Downward arrows indicate velocities of the five CN hyperfine lines.}
\label{spectra}
\end{figure}

Figure~\ref{spectra} shows examples of the CN, C$^{18}$O, and $^{13}$CO spectra extracted from local peaks in DR21, DR18, and DR17, where the CN intensities are the strongest in the observed area.
The intensity ratios of C$^{18}$O to $^{13}$CO are $\sim$1/6 in the local peaks.
Line ratios of CN to C$^{18}$O and the CN hyperfine lines are different from region to region, suggesting that optical depths and/or excitation states are different among the regions.
The theoretical CN hyperfine line intensities relative to that of the strongest line are 0.37, 0.30, and 0.30 for $F$=3/2--1/2, $F$=1/2--1/2, and $F$=3/2--3/2 transitions, respectively, assuming the local thermodynamic equilibrium and the optically thin limit (\citealt{Skatrud83}).
Among the regions, hyperfine line ratios in DR21 are larger than the theoretical values (0.57$\pm$0.05, 0.42$\pm$0.04, and 0.44$\pm$0.04 for $F$=3/2--1/2, $F$=1/2--1/2, and $F$=3/2--3/2 transitions, respectively), while those in DR17 and DR18 are roughly comparable to the theoretical values.
Therefore, CN may be optically thick in the local peak of DR21.
Unfortunately, the noise level of the CN data is too large to discuss the hyperfine line ratios except for strong local peaks.
The fraction of regions where the CN hyperfine line ratio has S/N$>$3 is only $\sim$10\% of the whole region.
Therefore, we use only the strongest CN line in the following discussion, and do not discuss variations of hyperfine line ratios and optical depth derived from the ratios.

\begin{figure}
\epsscale{0.9}
\plotone{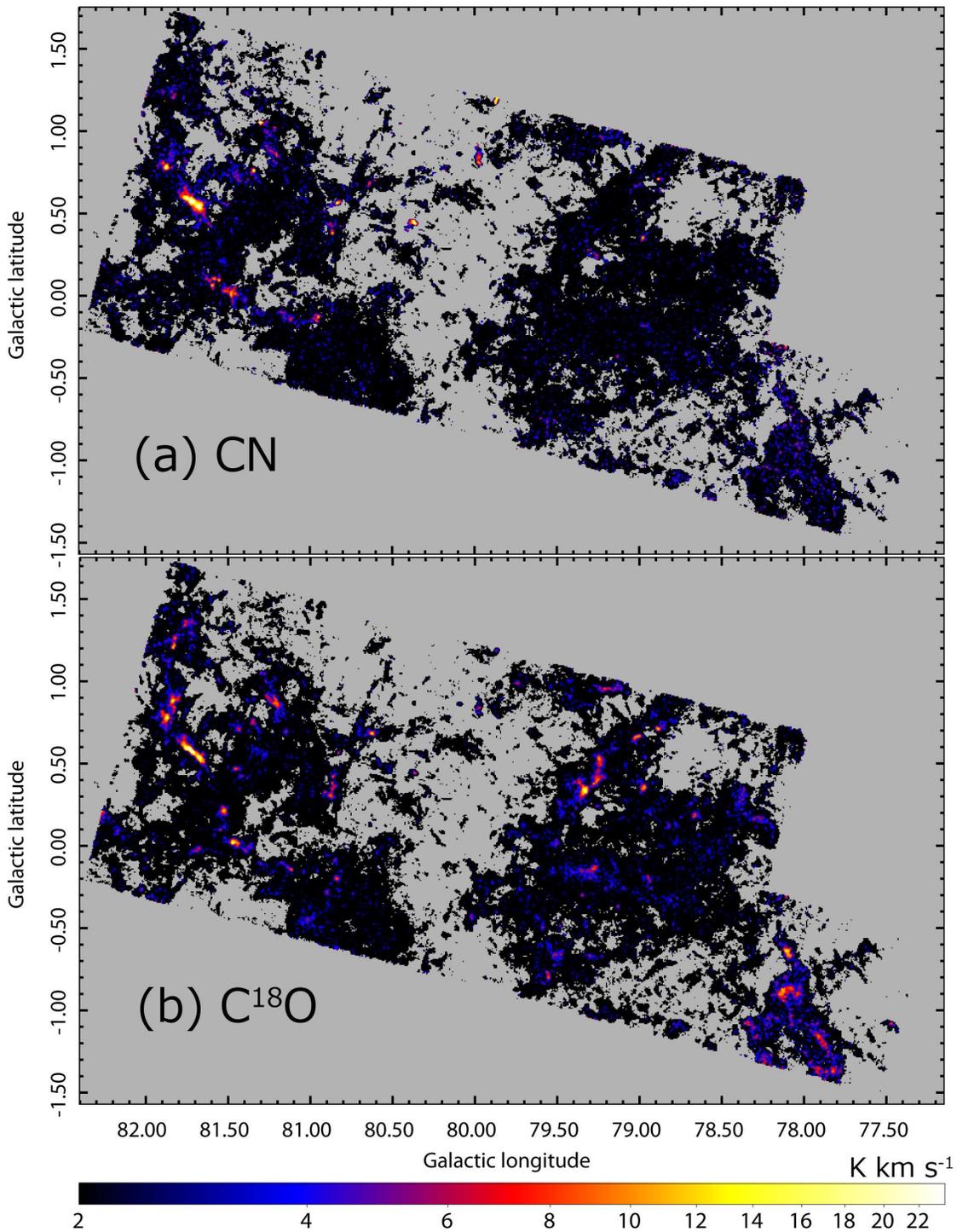}
\caption{Integrated intensity maps of (a) CN ($N$=1--0, $J$=3/2--1/2, $F$=5/2--3/2) and (b) C$^{18}$O ($J$=1--0). Regions where $^{13}$CO emission is not significantly detected are masked out.}
\label{integrated}
\end{figure}

Figure~\ref{integrated} shows integrated intensity maps of CN and C$^{18}$O, where emission in a range of $\pm$3~km~s$^{-1}$ relative to the systemic velocity is integrated.
Systemic velocity in each position is determined from the peak of the corresponding $^{13}$CO spectrum obtained in the present survey because $^{13}$CO, C$^{18}$O, and CN have the mostly same velocity (Fig.~\ref{spectra}).
Many spatial structures are clearly recognized in the maps.
It should be noted that the spatial extent of the CN and C$^{18}$O maps is comparable.
The critical density of C$^{18}$O is $\sim$10$^3$~cm$^{-3}$ (\citealt{Schoier05}), which is consistent with those estimated from observations (\citealt{Ikeda09}).
In addition, the effective critical density of CN is $\sim$10$^{3-4}$~cm$^{-3}$ (\citealt{Shirley15}).
Therefore, the similarity on the spatial extent suggests that both CN and C$^{18}$O are emitted from regions with the density of $\sim$10$^{3-4}$~cm$^{-3}$.
Notice that C$^{18}$O is widely distributed in both the North and South, while CN is rather concentrated in the North.
The spatial distribution of C$^{18}$O indicates that dense molecular gas is widely present in the North and South.
A remarkable difference between the North and South is the star-formation activity; the North shows more active star formation than the South (\citealt{Schneider06}).
These results suggest that the CN abundance is not simply proportional to the column density of molecular gas, and is related to the star-formation activity.

\begin{figure}
\epsscale{.75}
\plotone{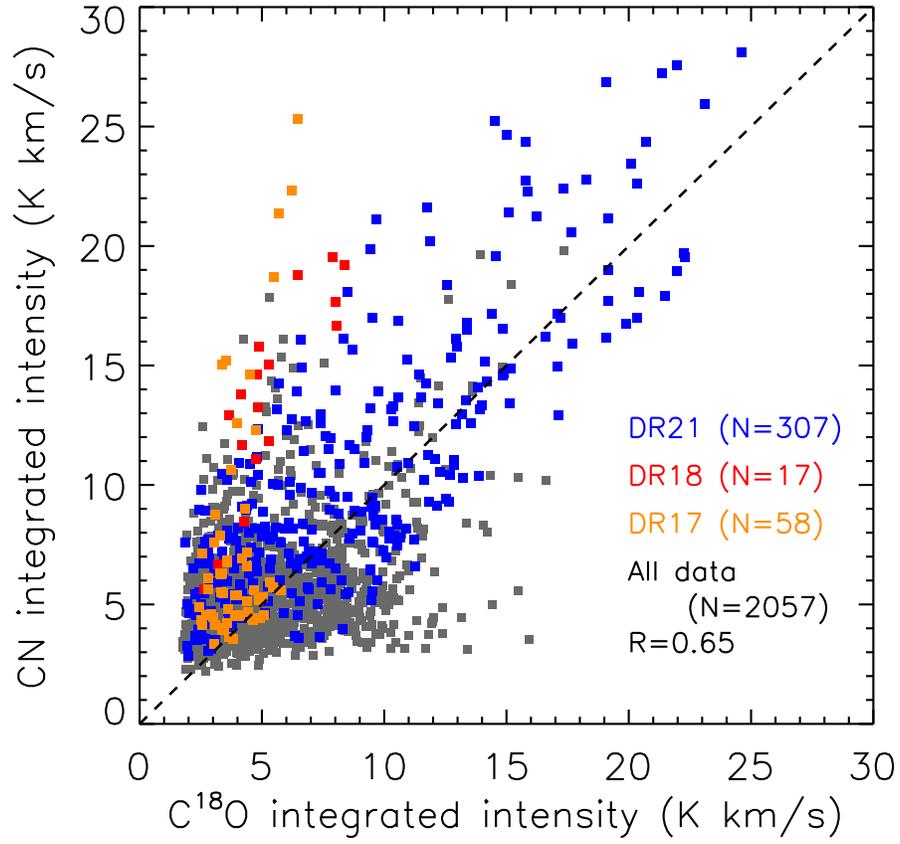}
\caption{Correlation between the integrated intensities of CN and C$^{18}$O. Regions where both CN and C$^{18}$O are significantly detected (S/N$>$3) are plotted. Linear correlation coefficient for all the data and the number of data points are shown in the bottom right.}
\label{correlation}
\end{figure}

Figure~\ref{correlation} shows correlation between the CN and C$^{18}$O integrated intensities, where there is a significant correlation as a whole ($R$=+0.65, $N$=2057).
There is, however, a large variation in the CN/C$^{18}$O integrated intensity ratios.
In order to identify the origin of the large variation, data points in the three CN intense regions (DR21, DR18, and DR17) are color coded.
It is clear that CN/C$^{18}$O ratios are systematically different among the CN intense regions, and that CN/C$^{18}$O ratios are especially enhanced in DR18 and DR17.
Additionally, CN/C$^{18}$O ratios are not likely to be uniform even within a region; CN/C$^{18}$O ratios significantly vary, ranging from 0.5 to 3 for DR21, from 2 to 3 for DR18, and from 1 to 5 for DR17.
In order to show significance of the variations, we fit a linear function to the data in Fig.~\ref{correlation}, and obtain reduced chi-square of 5.0, 0.64, 2.2 for DR21, DR18, and DR17, respectively, which indicates that CN/C$^{18}$O ratios are not uniform in DR21 and DR17 with the 99\% confidence level.
These results suggest that CN/C$^{18}$O ratios are strongly affected by the differences in the local interstellar environment.

The position showing intense CN emission in DR17 corresponds to one of the DR17 pillars identified by \citet{Schneider06} using KOSMA $^{13}$CO (2--1) data.
They suggested that the pillar structures are locally shaped by OB stars in DR17.
\citet{Gottschalk12} also identified the same structure using JCMT $^{12}$CO (3--2) data, and showed that the cometary structure points to the stellar cluster Cl 14, which contains 12 OB stars.
In addition, \citet{Schneider06} detected a molecular clump in DR18.
They argued that the clump is affected by the Cygnus OB2 cluster since it has an illuminated tip facing the cluster and a tail pointing away from its center.
\citet{Comeron99} found that the molecular clump in DR18 is also illuminated by a nearby B0.5V type star.
As already discussed in these studies, DR17 and DR18 have cometary structures which are likely to be formed by the intense UV radiation from the nearby OB stars.
Hence the enhancement of CN/C$^{18}$O ratios may be caused by the difference in the UV radiation environment.
A part of DR21 also shows high CN/C$^{18}$O ratios comparable with DR17 and DR18, although DR21 does not show such clear cometary structures.
Considering a number of widely distributed massive stars and YSOs around DR21 (\citealt{Beerer10}), it is possible that intense interstellar UV radiation due to nearby massive stars enhances the CN/C$^{18}$O ratios.
Alternatively, deeply embedded massive stars may locally enhance the CN/C$^{18}$O ratios from inside the molecular cloud.

%%-------------------------------------------------------------------------------------------------------------------------------------------------------
\section{Discussion}

\begin{figure}
\epsscale{0.75}
\plotone{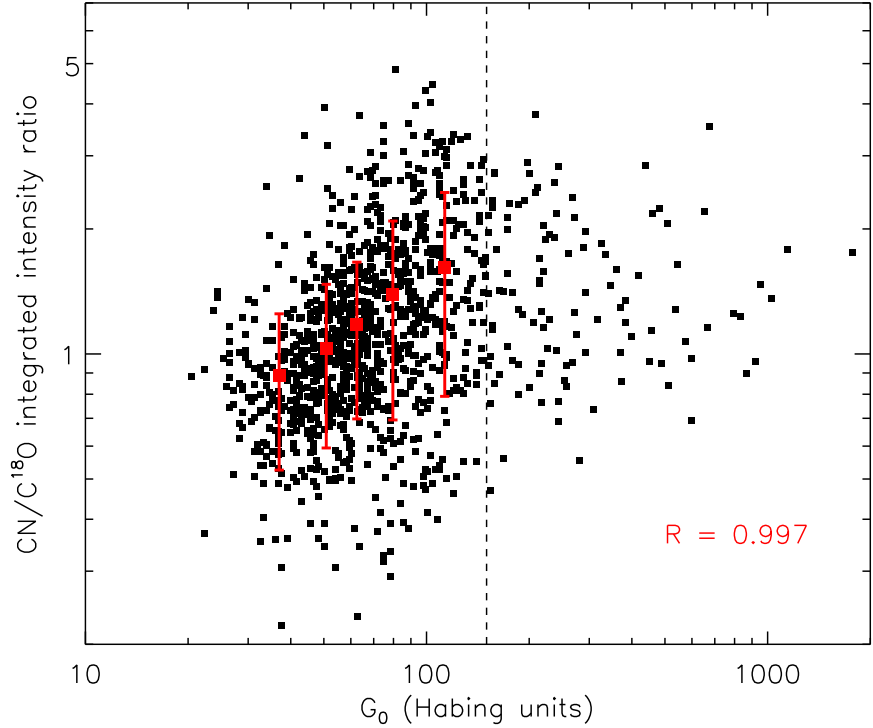}
\caption{CN/C$^{18}$O integrated intensity ratios plotted against $G_0$ for regions where CN/C$^{18}$O ratios are detected with S/N$>$3. The red data points show averages calculated for five $G_0$ ranges in $G_0<150$ (dashed line), where each $G_0$ range includes $\sim$220 data points. The red errorbar indicates 1$\sigma$ standard deviation in each $G_0$ range. The correlation coefficient for the red data is shown in the bottom right.}
\label{ratiodust}
\end{figure}

Connection between the CN/C$^{18}$O ratio and the UV radiation is suggested in the previous section.
We quantify the strength of the UV radiation, and directly examine the relation between the CN/C$^{18}$O ratio and the UV radiation.
As a probe of the UV radiation, we use the far-UV intensity normalized to that of the solar neighborhood, $G_0$, which is estimated using dust temperature ($T_{\mathrm{dust}}$): $G_0 = (T_\mathrm{dust}/12.2~\mathrm{K})^5$ (\citealt{Hollenbach91}).
\citet{Hollenbach91} assumed the typical density in PDRs of $10^{2-5}$~cm$^{-3}$.
As discussed in the previous section, both C$^{18}$O and CN are likely to be originated from regions with $\sim10^{3-4}
$~cm$^{-3}$, which is comparable with the typical density.
Therefore, the equation is applicable throughout the observed area.
The dust temperature was derived from the color of the $Herschel$/PACS 70~$\micron$ and 160~$\micron$ maps which were retrieved from NASA/IPAC Infrared Science Archives, assuming the power-law dust emissivity index of $\beta=1$.
The background emission was estimated from a region ($l$=80.88$^\circ$, $b$=2.47$^\circ$), and subtracted from the maps.
The 70~$\micron$ and 160~$\micron$ bands are suitable for deriving temperature of the warm dust component heated by the UV radiation.
Since warm dust shows small $\beta$ (e.g., \citealt{Rodon10}), we assumed $\beta$=1.
The temperature estimated in \citet{Schneider16} is lower than that in the present study by 10--15~K, which is caused by differences in the photometric bands and $\beta$ (160--500~$\micron$ maps; $\beta$=2). 
Figure~\ref{ratiodust} shows the CN/C$^{18}$O ratios plotted against $G_0$.
We find that CN/C$^{18}$O ratios significantly correlate with $G_0$ ($R$=+0.997, $N$=5) in $G_0<$150, indicating that the CN/C$^{18}$O ratios are enhanced by the intense UV radiation via photodissociation reactions.
This result also indicates that the CN/C$^{18}$O ratio is a good probe of PDRs.
In regions with $G_0>$150, CN/C$^{18}$O ratios are not likely to follow the relation determined in $G_0<$150.
One possible cause of the discrepancy is the overestimate of $G_0$; since one-component modified blackbody is assumed to estimate $T_{\mathrm{dust}}$, $G_0$ is overestimated if emission from even warmer component and/or stochastic heating contaminates the 70~$\micron$ map.
Another possibility is the underestimate of CN/C$^{18}$O ratios; CN might be photo-dissociated due to the UV radiation (\citealt{Boger05}).

\begin{figure}
\epsscale{.75}
\plotone{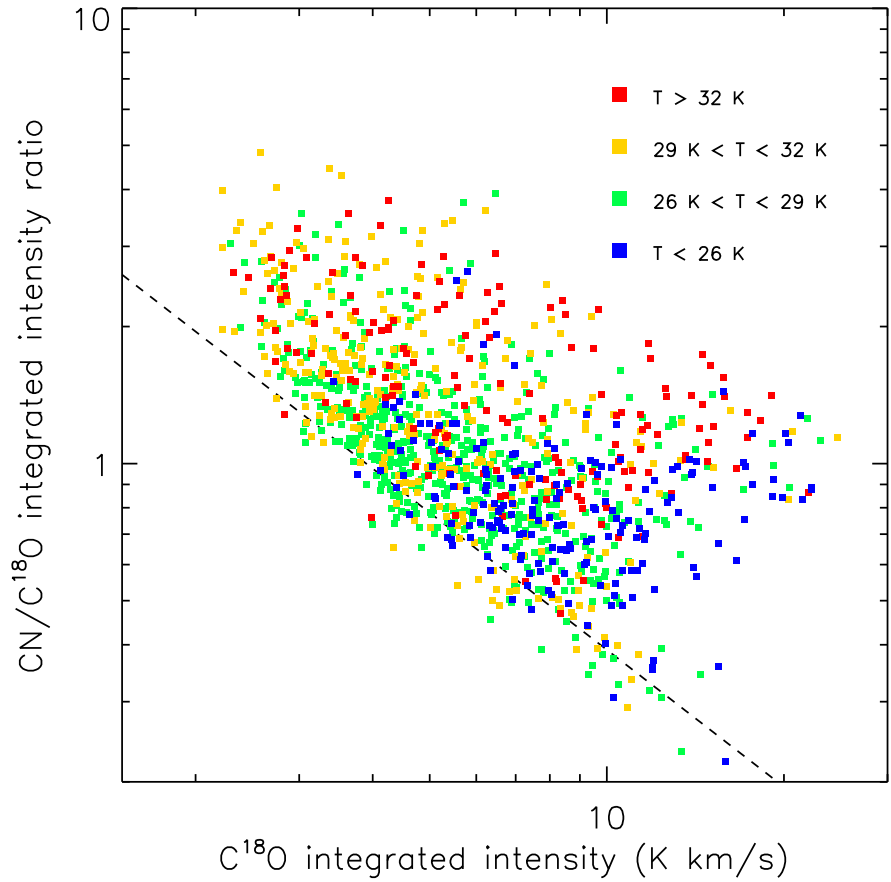}
\caption{CN/C$^{18}$O integrated intensity ratios plotted against C$^{18}$O integrated intensities for regions where CN/C$^{18}$O ratios are detected with S/N$>$3. The data are color-coded into four levels according to the dust temperature. The dashed line indicates the typical 3$\sigma$ detection limit of CN.}
\label{ratio}
\end{figure}

Next, we examine variations of CN/C$^{18}$O ratios inside molecular clouds.
Figure~\ref{ratio} shows CN/C$^{18}$O ratios as a function of C$^{18}$O integrated intensities, which roughly correspond to $A_\mathrm{V}$.
As seen in the figure, there is a declining trend between CN/C$^{18}$O ratios and C$^{18}$O integrated intensities.
Note that there are no data points in the lower left corner due to an observational bias (i.e., detection limit of CN).
We, however, argue that the declining trend is real because there are also no data points in the upper right corner.
The overall declining trend in Fig.~\ref{ratio} is interpreted as a simple situation: a molecular cloud illuminated by the interstellar (external) UV radiation from the nearby massive stars.
In this situation, photodissociation reactions are expected to be active at the surface of a molecular cloud (i.e., small C$^{18}$O integrated intensity), while they are not inside a molecular cloud (i.e., large C$^{18}$O integrated intensity) because the UV radiation is reduced due to the interstellar extinction.
Therefore, the regions with high and low CN/C$^{18}$O ratios may correspond to the surface and inside of the molecular clouds, respectively.
In order to support this idea, the data points in Fig.~\ref{ratio} are color-coded according to the dust temperature because the dust temperature is also available as a probe of the position in a cloud; dust temperature decrease from the surface to inside of a molecular cloud in the simple situation.
The median of CN/C$^{18}$O ratios are 0.81, 1.03, 1.34, and 1.41 for regions with $T_{\mathrm{dust}}<$26~K, 26~K$<T_{\mathrm{dust}}<$29~K, 29~K$<T_{\mathrm{dust}}<$32~K, and $T_{\mathrm{dust}}>$32~K, respectively.
The CN/C$^{18}$O ratio increases  with the dust temperature, which is also consistent with the simple situation.
Hence the variations in the CN/C$^{18}$O ratio as a function of the C$^{18}$O integrated intensity and the dust temperature suggest that CN/C$^{18}$O ratios are high at surfaces of molecular clouds illuminated by the interstellar UV radiation, while they are low inside molecular clouds due to the weak UV radiation.

\begin{figure}
\epsscale{0.75}
\plotone{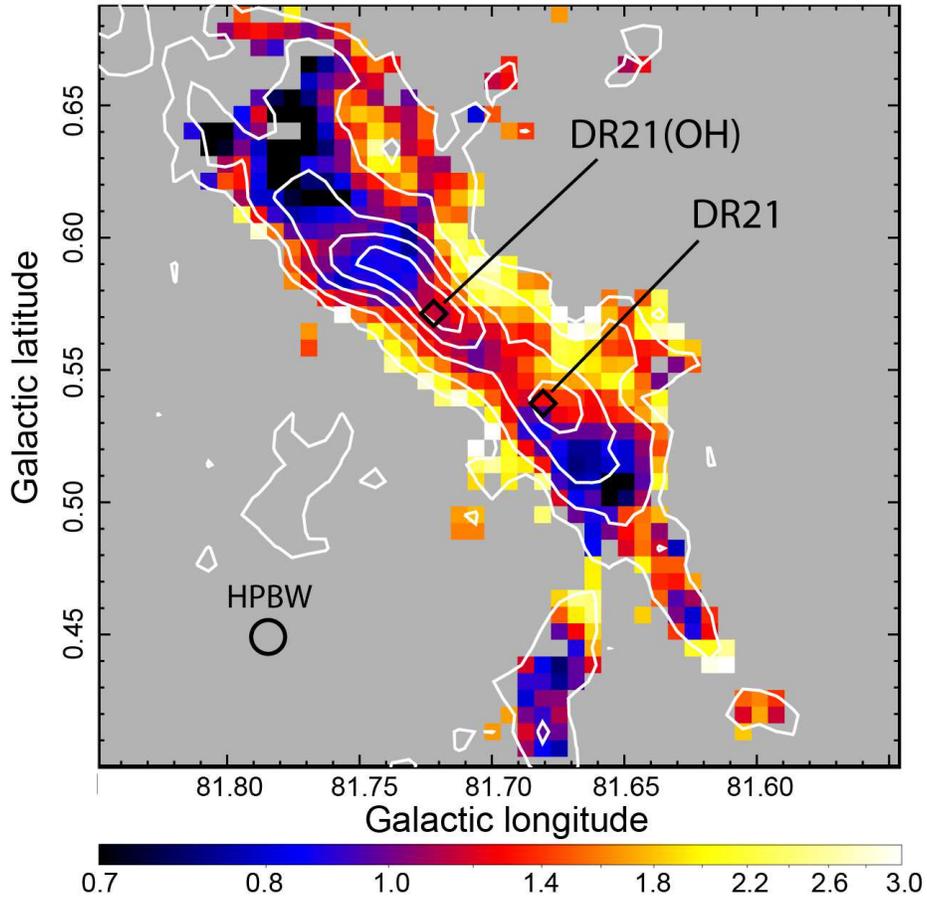}
\caption{CN/C$^{18}$O integrated intensity ratio map of DR21. Contours indicate C$^{18}$O integrated intensity at 2.5, 5, 10, 15, 20~K~km~s$^{-1}$. Diamonds indicate positions of the star-forming sites, DR21 and DR21(OH).}
\label{ratiomap}
\end{figure}

We finally check variations of CN/C$^{18}$O ratios on the local map.
Since most of the CN and C$^{18}$O detected regions are too small to examine the inner structures with the beam in the present study (46$\arcsec$), we focus on the DR21 region which is the largest region where CN and C$^{18}$O are detected (Fig.~\ref{integrated}).
Figure~\ref{ratiomap} shows the CN/C$^{18}$O ratio map of DR21.
In the map, CN/C$^{18}$O ratios are high in the outer part of the molecular cloud, while they are low inside the cloud; the median of CN/C$^{18}$O ratios at the outer (C$^{18}$O integrated intensity $<$5~K~km~s$^{-1}$; 2nd contour in Fig.~\ref{ratiomap}) and inner ($>$5~K~km~s$^{-1}$) parts are 1.90 and 1.15, respectively.
This variation is consistent with Fig.~\ref{ratio}, and supports that photodissociation reactions due to the interstellar UV radiation is dominant at the outer part of DR21.
In addition, we find that CN/C$^{18}$O ratios are locally enhanced near the star-forming sites (DR21 and DR21(OH)), suggesting that contribution of UV emission from embedded massive stars is not negligible.
However, since CN/C$^{18}$O ratios at the positions of DR21 and DR21(OH) ($\sim$1.2) are smaller than that at the surface (1.90), the UV emission from embedded massive stars may not be the dominant factor to affect CN/C$^{18}$O ratios in DR21.
These map-based results confirm that variations of CN/C$^{18}$O ratios are consistent with the characteristics of the interaction between the ISM and UV.
Hence we conclude that the CN/C$^{18}$O ratio is controlled by the UV radiation, and is available as a good probe of PDRs.

%%-------------------------------------------------------------------------------------------------------------------------------------------------------
\section{Conclusion}

We have tested the availability of the CN/C$^{18}$O ratio as a PDR probe for Cygnus-X based on unbiased large scale (9~deg$^2$) CN and C$^{18}$O maps obtained in the Nobeyama 45m Cygnus-X CO survey.
CN and C$^{18}$O are detected in a variety of objects in the Cygnus-X North and South.
C$^{18}$O is widely distributed in both the Cygnus-X North and South, while CN is rather concentrated in the North, suggesting that the star formation activity is related to the CN formation.
We find that there is a significant correlation between CN and C$^{18}$O integrated intensities, although there is a large variation in CN/C$^{18}$O ratios.
We also find that the CN/C$^{18}$O ratios are systematically different from region to region, and they are especially enhanced in DR17 and DR18 which are irradiated by the nearby OB stars.
This result suggests that CN/C$^{18}$O ratios are enhanced by the intense UV radiation via photodissociation reactions.
Based on the results, we investigate the relation between the CN/C$^{18}$O ratio and strength of the UV radiation, and find that CN/C$^{18}$O ratios positively correlate with $G_0$.
We also find that CN/C$^{18}$O ratios tend to decrease inside molecular clouds, where the interstellar UV radiation is reduced due to the interstellar extinction.
Variations of CN/C$^{18}$O ratios are consistent with the characteristics of the interaction between the ISM and UV.
Hence we conclude that the CN/C$^{18}$O ratio is controlled by the UV radiation, and is available as a good probe of PDRs.

%%-------------------------------------------------------------------------------------------------------------------------------------------------------
\acknowledgments

We express many thanks to the anonymous referee for useful comments.
This work is based on observations with the 45m telescope in the Nobeyama Radio Observatory (NRO).
NRO is a branch of the National Astronomical Observatory of Japan, National Institutes of Natural Sciences.
This work is also based on archival data obtained with the Herschel Space Observatory (Observation IDs: 1342244168, 1342257386, 1342196917, 1342244170, 1342211307, 1342257384).
Data analysis was carried out on the open use data analysis computer system at the Astronomy Data Center, ADC, of the National Astronomical Observatory of Japan.
This research is supported by JSPS KAKENHI Grant Number JP17842380.

\bibliographystyle{aasjournal}

\end{document}